\documentclass[fleqn,usenatbib]{mnras}

\usepackage[dvipdfmx]{graphicx}

\usepackage{amsmath}
\usepackage{color}

\usepackage[T1]{fontenc}
\usepackage{aecompl}
\usepackage{yfonts}
\usepackage{longtable}

\newcommand{\lesssim}{\,\raisebox{-0.4ex}{$\stackrel{<}{\scriptstyle\sim}$}\,}
\newcommand{\gtrsim}{\,\raisebox{-0.4ex}{$\stackrel{>}{\scriptstyle\sim}$}\,}

\begin{document}

\title[]{The internal rotation of globular clusters revealed by Gaia DR2}

 \author[P. Bianchini et al.]{P. Bianchini$^{1,}$\thanks{E-mail:
bianchip@mcmaster.ca}\thanks{CITA National Fellow},
R. P. van der Marel$^{2,3}$,
A. del Pino$^{2}$,
L. L. Watkins$^{2}$,
\newauthor
A. Bellini$^{2}$,
M. A. Fardal$^{2}$,
M. Libralato$^{2}$, \&
A. Sills$^{1}$
\\
$^{1}$Department of Physics and Astronomy, McMaster University, Hamilton, ON, L8S 4M1, Canada\\
$^{2}$Space Telescope Science Institute, 3700 San Martin Drive, Baltimore MD 21218, USA\\
$^{3}$Center for Astrophysical Sciences, Department of Physics \& Astronomy, Johns Hopkins University, Baltimore, MD 21218, USA
}

\date{}
\maketitle

\begin{abstract}
Line-of-sight kinematic studies indicate that many Galactic globular clusters have a significant degree of internal rotation. However, three-dimensional kinematics from a combination of proper motions and line-of-sight velocities are needed to unveil the role of angular momentum in the formation and evolution of these old stellar systems. Here we present the first quantitative study of internal rotation on the plane-of-the-sky for a large sample of globular clusters using proper motions from \textit{Gaia} DR2. We detect signatures of rotation in the tangential component of proper motions for 11 out of 51 clusters at a $>$3-sigma confidence level, confirming the detection reported in \citet{Helmi2018} for 8 clusters, and additionally identify 11 GCs with a 2-sigma rotation detection. For the clusters with a detected global rotation, we construct the two-dimensional rotation maps and proper motion rotation curves, and we assess the relevance of rotation with respect to random motions ($V/\sigma\sim0.08-0.51$). We find evidence of a correlation between the degree of internal rotation and relaxation time, highlighting the importance of long-term dynamical evolution in shaping the clusters current properties. This is a strong indication that angular momentum must have played a fundamental role in the earliest phases of cluster formation. Finally, exploiting the spatial information of the rotation maps and a comparison with line-of-sight data, we provide an estimate of the inclination of the rotation axis for a subset of 8 clusters. Our work demonstrates the potential of \textit{Gaia} data for internal kinematic studies of globular clusters and provides the first step to reconstruct their intrinsic three-dimensional structure. 

\end{abstract}



\section{Introduction}


Globular clusters (GCs) are old stellar systems ($>10$ Gyr) shaped by the complex interplay between internal evolutionary processes (stellar evolution and dynamical evolution) and external processes (interaction with the host galaxy). The details of their formation and subsequent long-term evolution remain an unanswered question in modern astrophysics.

GCs have long been considered spherical, non-rotating stellar systems, given that internal relaxation processes would naturally dissipate angular momentum on a long-term evolutionary scale (see e.g. \citealp{Tiongco2017} and references therein). However, in recent years, an increasing number of line-of-sight kinematic studies have shown evidence that a significant amount of internal rotation is actually present in many present-day Milky Way (MW) GCs (\citealp{Ferraro2018,Lanzoni2018,Kamann2018,Boberg2017,Jeffreson2017,Lardo2015,Kimmig2015,Fabricius2014,Kacharov2014,Bianchini2013,Bellazzini2012,Lane2011}). As of today, more than 50\% of the GCs sampled show clear signatures of internal rotation (\citealp{Kamann2018}). Moreover, signatures of rotation have also been reported for intermediate-age clusters (e.g. \citealp{Mackey2013}), young massive clusters (\citealp{Henault-Brunet2012}) and nuclear star clusters (\citealp{Feldmeier2014,Nguyen2018}),  indicating that internal rotation is a common ingredient across dense stellar systems of different scales.  

On the theoretical side, the presence of a significant amount of internal rotation raises a series of fundamental issues connected to the formation and evolution of GCs. First, rotation is expected to change the long-term dynamical evolution of GCs, with a number of studies indicating that rotation accelerates the evolution (e.g. \citealp{EinselSpurzem1999,Kim2008,Hong2013}) and strongly shapes their present day morphology (\citealp{vandenBergh2008,Bianchini2013}). Moreover, even a low amount of angular momentum in todays GCs could represent the fossil-record of strong primordial rotation in proto-GCs (\citealp{Tiongco2017}, \citealp{Mapelli2017}) or indicate a peculiar evolution environment (\citealp{Vesperini2014,Tiongco2018}). Finally, signatures of internal rotation could be crucial in revealing the formation mechanisms of the still-unsolved puzzle of multiple stellar populations in GCs (\citealp{Mastrobuono-battisti2013,Henault-Brunet2015,MastrobuonoBattisti2016,Cordero2017}).

Despite the recent advances in the study of internal rotation along the line-of-sight, only a handful of works have provided direct measurements of rotation on the plane-of-the-sky. In fact, high accuracy \textit{Hubble Space Telescope} (\textit{HST}) proper motions (PMs) only deliver relative measurements, implying that any solid body rotation is canceled out in the data due to the astrometric reduction (\citealp{McNamara2003,McLaughlin2006,Anderson2010,Bellini2014,Watkins2015,Watkins2015b}). A few exceptions include the study of NGC5139 by \citet{vanLeeuwen2000} with photographic plates spanning several decades; the \textit{HST} studies of NGC104 (\citealp{Anderson2003,Bellini2017}) and NGC362 (\citealp{Libralato2018}) exploiting the background stars of the Small Magellanic Cloud as an absolute reference frame against which to measure the rotation of the cluster; and the \textit{HST} study of NGC6681 by \citet{Massari2013} using background stars of the Sagittarius dwarf spheroidal galaxy as a reference. Finally, indirect measurements of rotation on the plane-of-the-sky have been reported studying the shape of the proper motion distributions of NGC6656, NGC104 and NGC5139 (\citealp{Peterson1994, Heyl2017, Bellini2018}, respectively), or by geometric reconstruction for NGC5139 and NGC7078 (see \citealp{vandeVen2006} and \citealp{vandenBosch2006}, respectively).

In order to understand the role of angular momentum in GCs, a simultaneous analysis of rotation along the line-of-sight and on the plane-of-the-sky is necessary. This was demonstrated by \citet{Bellini2017} who discovered the presence of an unexpectedly high amount of intrinsic internal rotation in NGC104, with the combination of three-dimensional kinematics and dynamical models. \textit{Gaia} DR2 proper motions provide the optimal tool to directly detect internal rotation on the plane-of-the-sky (\citealp{Prusti2016,Brown2018,Helmi2018}) and the necessary first step for a three-dimensional analysis of GC dynamics. \citet{Helmi2018} demonstrated the feasibility of these measurements with DR2 and reported a qualitative detection of rotation for 8 GCs. Moreover, an analysis of the rotation profiles of the different stellar populations of NGC104 was reported in \citet{Milone2018}. The goal of our paper is to provide a homogenous investigation of rotation on the plane-of-the-sky for a large sample of GCs, quantifying the amount of angular momentum and its spatial distribution, together with an interpretation of the three-dimensional structure of a GC. In Section 2 we describe the selection of our targets in \textit{Gaia} DR2, in Section 3 we present the procedure followed to clean the datasets and the analysis to measure the global rotation, the proper motion rotation profile and rotation maps. In Section 4 we present our results, we assess the role of angular momentum with respect to random motions, and estimate the inclination angle of the rotation axis for a subsample of clusters. Finally, in Section 5 we discuss our conclusions.

\section{Target selection}
\label{sec:2}
We consider a total of 85 Milky Way globular clusters with distance from the Sun $<15$ kpc. For these clusters, we perform mock observations of PM rotation in order to determine for which systems a rotation signal could be detectable with \textit{Gaia} DR2.

For each cluster we select all stars within three half-light radii ($3R_h$, with $R_h$ and the coordinates of the centre of the clusters from the Harris Catalogue, \citealp{Harris1996}, 2010 edition) from \textit{Gaia} Data Release 1 (\citealp{Prusti2016,Brown2016}). For each of the stars we assign a mock velocity field consisting of a radially-constant velocity dispersion for both the radial and tangential velocity component $\mu_r$ and $\mu_t$, drawn from a Gaussian distribution centred on zero and with a dispersion of 10 km s$^{-1}$ . Moreover, we add a rotational component in the tangential direction given by
\begin{equation}
\mu_t=\frac{2V_{peak}}{R_{peak}}\frac{R}{1+(R/R_{peak})^2},
\label{eq:rotation_profile}
\end{equation}
describing the typical differential rotation profiles of GCs observed in line-of-sight velocity measurements (e.g. \citealp{Kacharov2014,Mackey2013}). We consider as the peak of the velocity profile $V_{peak}=1$ km s$^{-1}$ and as the radial position of the rotation peak $R_{peak}=R_h$. Finally, we assign to each velocity an estimated measurement error computed using the PyGaia package,\footnote{\url{https://github.com/agabrown/PyGaia}} in order to reproduce the expected proper motion accuracy of \textit{Gaia} DR2. Our mock observations assume that all the stars measured within $3R_h$ are genuine cluster members; moreover we restrict our analysis to within $3R_h$ from the cluster centres since this is the typical region where the rotation signal is expected to be significant (e.g. \citealp{Bianchini2013,Ferraro2018}).

For each cluster, we measure the rotation signal  (i.e. the mean value of the $\mu_t$ velocity component) and the associated 68\% uncertainties considering the log-likelihood function
\begin{equation}
-\log \lambda=-\log \prod^{N}_{i=1} p(v_i,\epsilon_i)=-\sum^{N}_{i=1}\log p(v_i,\epsilon_i),
\label{eq:loglike}
\end{equation}
with $p(v_i,\epsilon_i)$ the probability of finding a star with velocity $v_i$ and uncertainty $\epsilon_i$, given a mean velocity $v$ and velocity dispersion $\sigma$ (see e.g. \citealp{Kamann2018})
\begin{equation}
p(v_i,\epsilon_i)=\frac{1}{2\pi\sqrt{\sigma^2+\epsilon_i^2}}\exp{ -\frac{(v_i-v)^2}{2(\sigma^2+\epsilon_i^2)}}.
\label{likelihood}
\end{equation}
We compute the average rotation in the tangential component within $3R_h$, the velocity dispersion and the associated 1-sigma errors sampling the log-likelihood function (equation \ref{eq:loglike}) using the Markov Chain Monte Carlo algorithm \textit{emcee} by \citet{Foreman-Mackey2013}. 
Clusters for which the rotation signal was recovered with at least a 3-sigma confidence level were selected as our targets and further considered for the analysis of the \textit{Gaia} DR2 dataset. The GCs that did not pass the cuts are typically characterized by numbers of stars in the field-of-view $N\lesssim10^3$ and are at distances $d\gtrsim10$ kpc. The list of the selected 51 GCs are reported in Table \ref{tab:1}. These clusters are the objects for which we should be able to detect at least a 1 km s$^{-1}$ rotation signal given their distance and number of member stars.

\begin{table*}
\tabcolsep=0.10cm
\begin{center}
\caption{Properties of the 51 selected clusters and their derived kinematic properties from \textit{Gaia} DR2. For all GCs we report the distance $d$, the half-light radius R$_h$, the metallicity [Fe/H], distance modulus (m$-$M)$_V$ and reddening E$_{(B-V)}$ from the Harris catalog (\citealp{Harris1996}), the central velocity dispersion $\sigma_0$ from \citet{Baumgardt2018}, the number of selected member stars from \textit{Gaia} DR2, and the measured rotation signal in the tangential proper motion component within 3 half-light radii $\mu_t(<3R_h)$ from this work. For clusters with a 3-sigma rotation detection, we further report the value and the position of the peak of the rotation profile, $V_{peak}$ and $R_{peak}$ and the $V/\sigma$ parameter, defined as the ratio of the proper motion rotation peak and the central velocity dispersion. For all the other clusters we report the 1-sigma upper limit of the $V/\sigma$ parameter.}

\begin{tabular}{lcccccclrcccc}
\hline\hline
& d & R$_h$ & [Fe/H] & (m$-$M)$_V$& E$_{(B-V)}$ & $\sigma_0$ & N & $\mu_t(<3R_h)$ &\multicolumn{2}{c}{$V_{peak}$} & $R_{peak}/R_h$ & $V/\sigma$\\
& kpc & arcmin & & mag  & mag & km s$^{-1}$& & mas yr$^{-1}$ &mas yr$^{-1}$ & km s$^{-1}$  & & \\
\hline
NGC104 & 4.5 & 3.17 & -0.72 & 13.37 & 0.04 & 12.2 & 28327 & $-0.251^{+0.004}_{-0.005}$ &0.291$\pm{0.006}$ & -6.21$\pm{0.13}$ & 1.46$\pm{0.07}$ & $0.51\pm{0.01}$\\
NGC288 & 8.9 & 2.23 & -1.32 & 14.84 & 0.03 & 3.3 & 6838 & $0.014^{+0.012}_{-0.012}$ & & & & 0.33\\
NGC362 & 8.6 & 0.82 & -1.26 & 14.83 & 0.05 & 8.8 & 2191 & $-0.02^{+0.030}_{-0.028}$ & & & & 0.23\\
NGC1851 & 12.1 & 0.51 & -1.18 & 15.47 & 0.02 & 10.2 & 823 & $0.012^{+0.042}_{-0.043}$& & & & 0.30\\
NGC1904 & 12.9 & 0.65 & -1.6 & 15.59 & 0.01 & 6.5 & 1382 & $0.067^{+0.037}_{-0.038}$& & & &0.99\\
NGC2808 & 9.6 & 0.8 & -1.14 & 15.59 & 0.22 & 14.4 & 1835 & $-0.022^{+0.025}_{-0.023}$& & & &0.15\\
NGC3201 & 4.9 & 3.1 & -1.59 & 14.2 & 0.24 & 4.5 & 12939 & $0.014^{+0.006}_{-0.007}$& & & &0.10\\
NGC4372 & 5.8 & 3.91 & -2.17 & 15.03 & 0.39 & 4.9 & 6626 & $-0.033^{+0.005}_{-0.005}$ & -0.042$\pm{0.005}$ & -1.15$\pm{0.14}$ & 2.05$\pm{0.61}$ & $0.23\pm{0.01}$\\
NGC4590 & 10.3 & 1.51 & -2.23 & 15.21 & 0.05 & 3.7 & 3431 & $-0.001^{+0.022}_{-0.023}$& & & &0.30\\
NGC4833 & 6.6 & 2.41 & -1.85 & 15.08 & 0.32 & 4.8 & 5933 & $0.009^{+0.008}_{-0.008}$& & & &0.11\\
NGC5139 & 5.2 & 5.0 & -1.53 & 13.94 & 0.12 & 17.6 & 37813 & $0.171^{+0.005}_{-0.005}$ & 0.251$\pm{0.017}$ & 6.19$\pm{0.42}$ & 0.95$\pm{0.1}$ & $0.35\pm{0.02}$\\
NGC5272 & 10.2 & 2.31 & -1.5 & 15.07 & 0.01 & 8.1 & 8678 & $-0.042^{+0.005}_{-0.005}$ & -0.043$\pm{0.004}$ & -2.08$\pm{0.19}$ & 1.81$\pm{0.58}$ & $0.26\pm{0.02}$\\
NGC5286 & 11.7 & 0.73 & -1.69 & 16.08 & 0.24 & 9.3 & 1115 & $-0.094^{+0.041}_{-0.042}$& & & &0.80\\
NGC5904 & 7.5 & 1.77 & -1.29 & 14.46 & 0.03 & 7.7 & 7321 & $0.096^{+0.010}_{-0.011}$ & 0.11$\pm{0.009}$ & 3.91$\pm{0.32}$ & 2.22$\pm{0.69}$ & $0.51\pm{0.05}$\\
NGC5927 & 7.7 & 1.1 & -0.49 & 15.82 & 0.45 & 6.5 & 1975 & $-0.028^{+0.020}_{-0.020}$& & & &0.27\\
NGC5946 & 10.6 & 0.89 & -1.29 & 16.79 & 0.54 & 4.0 & 556 & $0.038^{+0.034}_{-0.032}$& & & &0.90\\
NGC5986 & 10.4 & 0.98 & -1.59 & 15.96 & 0.28 & 8.3 & 2367 & $0.051^{+0.040}_{-0.040}$& & & &0.54\\
NGC6093 & 10.0 & 0.61 & -1.75 & 15.56 & 0.18 & 9.5 & 916 & $-0.012^{+0.040}_{-0.038}$& & & &0.26\\
NGC6121 & 2.2 & 4.33 & -1.16 & 12.82 & 0.35 & 4.6 & 13183 & $0.015^{+0.006}_{-0.005}$& & & &0.05\\
NGC6171 & 6.4 & 1.73 & -1.02 & 15.05 & 0.33 & 4.3 & 4731 & $0.025^{+0.019}_{-0.021}$& & & &0.31\\
NGC6205 & 7.1 & 1.69 & -1.53 & 14.33 & 0.02 & 9.2 & 9062 & $-0.035^{+0.014}_{-0.013}$& & & &0.18\\
NGC6218 & 4.8 & 1.77 & -1.37 & 14.01 & 0.19 & 4.5 & 8297 & $-0.002^{+0.012}_{-0.012}$& & & &0.07\\
NGC6254 & 4.4 & 1.95 & -1.56 & 14.08 & 0.28 & 6.2 & 8378 & $0.031^{+0.011}_{-0.013}$& & & &0.14\\
NGC6266 & 6.8 & 0.92 & -1.18 & 15.63 & 0.47 & 15.2 & 1238 & $0.061^{+0.023}_{-0.026}$& & & &0.18\\
NGC6273 & 8.8 & 1.32 & -1.74 & 15.9 & 0.38 & 11.0 & 2663 & $0.076^{+0.013}_{-0.015}$  & 0.076$\pm{0.01}$ & 3.17$\pm{0.42}$ & 1.52$\pm{0.61}$ & $0.29\pm{0.04}$\\
NGC6304 & 5.9 & 1.42 & -0.45 & 15.52 & 0.54 & 5.7 & 1047 & $-0.014^{+0.021}_{-0.022}$& & & &0.17\\
NGC6333 & 7.9 & 0.96 & -1.77 & 15.67 & 0.38 &   & 1584 & $-0.018^{+0.015}_{-0.016}$\\
NGC6341 & 8.3 & 1.02 & -2.31 & 14.65 & 0.02 & 8.0 & 3670 & $0.018^{+0.025}_{-0.024}$& & & &0.21\\
NGC6352 & 5.6 & 2.05 & -0.64 & 14.43 & 0.22 & 4.4 & 1707 & $0.001^{+0.014}_{-0.013}$& & & &0.09\\
NGC6366 & 3.5 & 2.92 & -0.59 & 14.94 & 0.71 & 3.0 & 4284 & $0.004^{+0.009}_{-0.009}$& & & &0.07\\
NGC6362 & 7.6 & 2.05 & -0.99 & 14.68 & 0.09 & 3.9 & 7243 & $0.006^{+0.008}_{-0.008}$& & & &0.12\\
NGC6388 & 9.9 & 0.52 & -0.55 & 16.13 & 0.37 & 18.2 & 612 & $0.028^{+0.036}_{-0.034}$& & & &0.16\\
NGC6402 & 9.3 & 1.3 & -1.28 & 16.69 & 0.6 & 11.1 & 2150 & $-0.031^{+0.011}_{-0.011}$& & & &0.16\\
NGC6401 & 10.6 & 1.91 & -1.02 & 17.35 & 0.72 &  & 399 & $0.015^{+0.019}_{-0.019}$\\
NGC6397 & 2.3 & 2.9 & -2.02 & 12.37 & 0.18 & 5.2 & 12031 & $-0.011^{+0.005}_{-0.005}$& & & &0.03\\
NGC6522 & 7.7 & 1.0 & -1.34 & 15.92 & 0.48 & 8.2 & 570 & $0.041^{+0.039}_{-0.04}$& & & &0.35\\
NGC6539 & 7.8 & 1.7 & -0.63 & 17.62 & 1.02 & 5.9 & 787 & $-0.032^{+0.015}_{-0.015}$& & & &0.29\\
NGC6544 & 3.0 & 1.21 & -1.4 & 14.71 & 0.76 & 6.4 & 1448 & $-0.007^{+0.019}_{-0.018}$& & & &0.06\\
NGC6541 & 7.5 & 1.06 & -1.81 & 14.82 & 0.14 & 8.7 & 1697 & $-0.032^{+0.013}_{-0.013}$& & & &0.18\\
NGC6626 & 5.5 & 1.97 & -1.32 & 14.95 & 0.4 & 12.6 & 2070 & $0.024^{+0.015}_{-0.014}$& & & &0.08\\
NGC6637 & 8.8 & 0.84 & -0.64 & 15.28 & 0.18 &  & 774 & $-0.021^{+0.019}_{-0.021}$\\
NGC6656 & 3.2 & 3.36 & -1.7 & 13.6 & 0.34 & 8.4 & 12473 & $0.129^{+0.005}_{-0.006}$ & 0.161$\pm{0.008}$ & 2.44$\pm{0.12}$ & 1.04$\pm{0.1}$ & $0.29\pm{0.03}$\\
NGC6681 & 9.0 & 0.71 & -1.62 & 14.99 & 0.07 & 7.1 & 1159 & $0.015^{+0.043}_{-0.039}$& & & &0.35\\
NGC6712 & 6.9 & 1.33 & -1.02 & 15.6 & 0.45 & 5.0 & 1011 & $-0.002^{+0.022}_{-0.021}$& & & &0.15\\
NGC6752 & 4.0 & 1.91 & -1.54 & 13.13 & 0.04 & 8.3 & 11038 & $-0.033^{+0.006}_{-0.006}$ & -0.037$\pm{0.004}$ & -0.7$\pm{0.08}$ & 1.49$\pm{0.41}$ & $0.08\pm{0.01}$\\
NGC6779 & 9.4 & 1.1 & -1.98 & 15.68 & 0.26 & 6.1 & 2211 & $-0.044^{+0.018}_{-0.017}$& & & &0.46\\
NGC6809 & 5.4 & 2.83 & -1.94 & 13.89 & 0.08 & 4.8 & 8316 & $0.037^{+0.005}_{-0.005}$ &0.055$\pm{0.017}$ & 1.41$\pm{0.44}$ & 3.46$\pm{1.82}$ & $0.29\pm{0.09}$\\
NGC6838 & 4.0 & 1.67 & -0.78 & 13.8 & 0.25 & 3.3 & 3493 & $-0.008^{+0.005}_{-0.005}$& & & &0.08\\
NGC7078 & 10.4 & 1.0 & -2.37 & 15.39 & 0.1 & 12.9 & 2211 & $0.088^{+0.012}_{-0.011}$ & 0.109$\pm{0.007}$ & 5.37$\pm{0.35}$ & 1.61$\pm{0.29}$ & $0.42\pm{0.04}$\\
NGC7089 & 11.5 & 1.06 & -1.65 & 15.5 & 0.06 & 10.6 & 1799 & $-0.055^{+0.013}_{-0.012}$ & -0.073$\pm{0.017}$ & -3.98$\pm{0.93}$ & 0.88$\pm{0.34}$ & $0.38\pm{0.01}$\\
NGC7099 & 8.1 & 1.03 & -2.27 & 14.64 & 0.03 & 5.5 & 2463 & $0.041^{+0.021}_{-0.022}$& & & &0.43\\

\hline

\end{tabular}
\label{tab:1}
\end{center}
\end{table*}

\section{Measurement of rotation on the plane of the sky}
For the 51 selected targets we consider all the stars from \textit{Gaia} DR2 within $3R_h$ of their centres. In this section we describe the procedure used to select cluster members, eliminate likely foreground and background contamination, and detect rotation signals.
\begin{figure*}
\centering
\includegraphics[width=.73\textwidth]{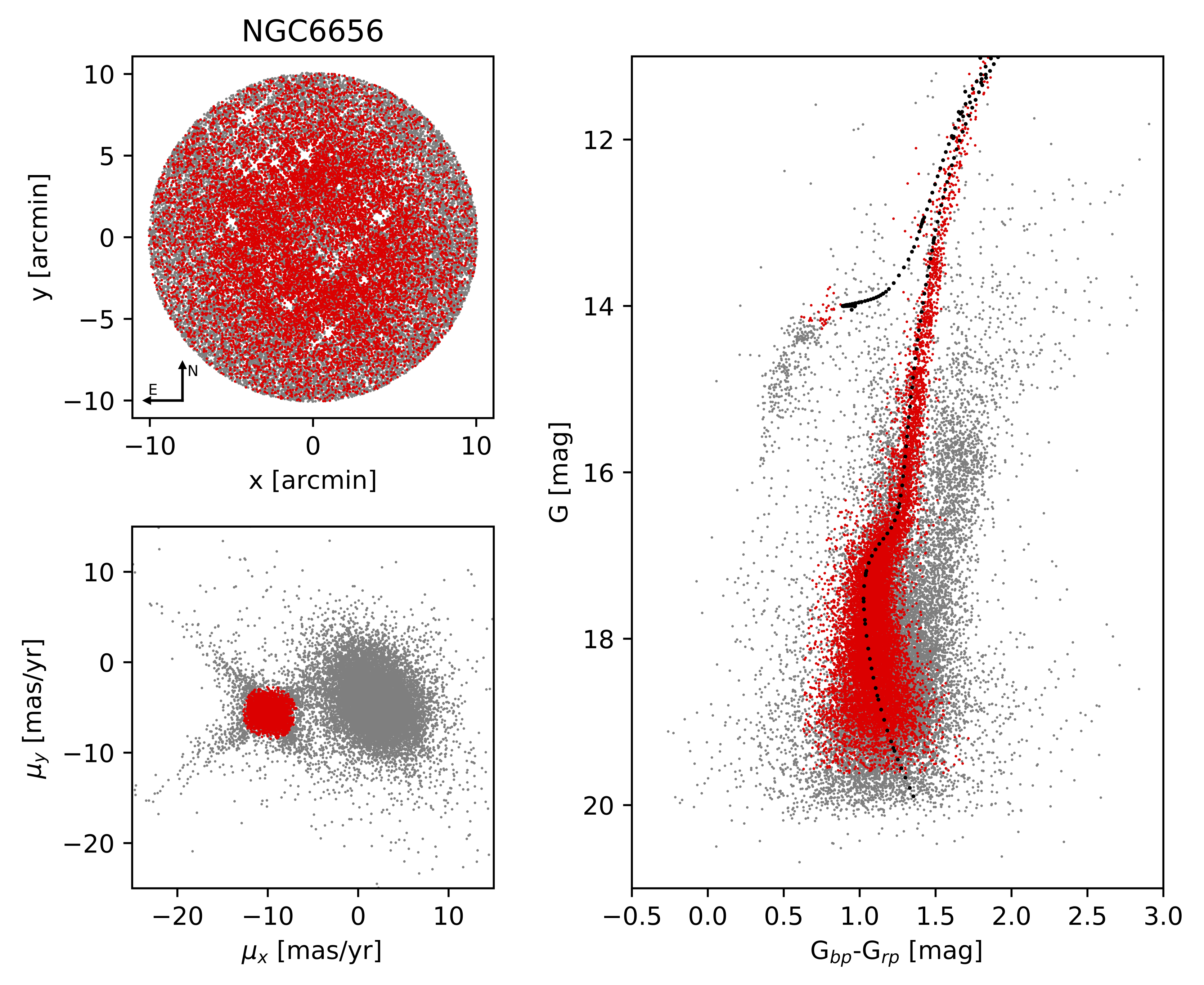}
\includegraphics[width=.73\textwidth]{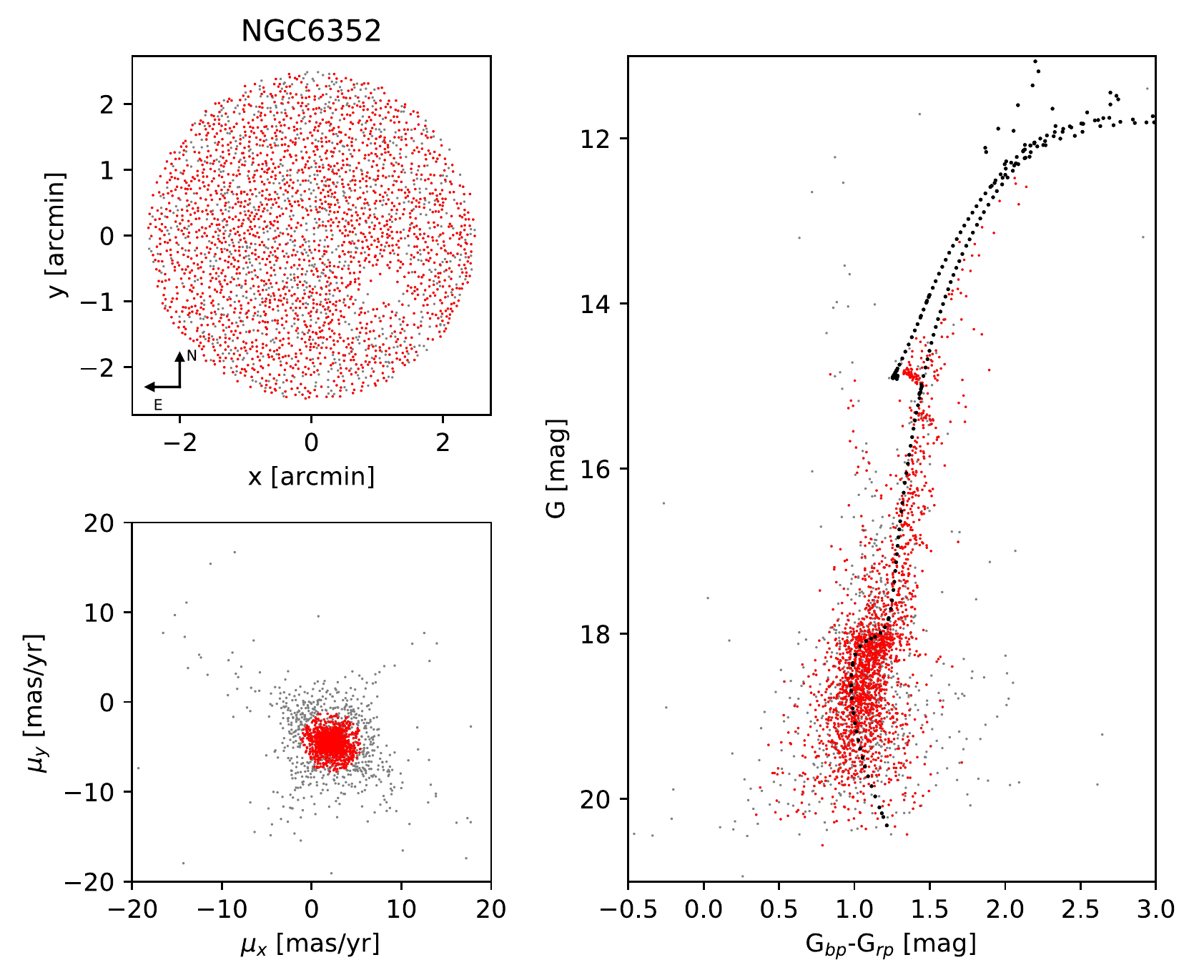}
\caption{Example of membership selection for NGC6656 and NGC6352. Red points indicate stars that passed all of the membership selection criteria (see Section \ref{sec:selection}), grey points are the stars excluded from our sample, black points the isochrones used for the CMD cleaning. \textit{Top left:} field-of-view within 3 half-light radii of the cluster centre. \textit{Bottom left:} proper motions along the x and y axis for the stars in the field-of-view. \textit{Right:} colour magnitude diagram for the stars in the field-of-view.}
\label{fig:selection}
\end{figure*}


\subsection{Selection of cluster members}
\label{sec:selection}

First, we clean each cluster sample of poorly measured stars based on their photometric and astrometric errors, here indicated as $\sigma$. All stars with photometric or astrometric errors larger than 3$\left < \sigma \right>$ at a given magnitude are rejected.
We next consider a set of nested criteria for our membership selection based on the stars' parallaxes (1), position in the colour-magnitude diagram CMD (2), and their PM (3).

In step (1), each list is cleaned of obvious MW contaminants based on their heliocentric distances inferred from parallaxes. Given the distance of a star $d$ and its error $\sigma_d$, stars at distances of $d + 3\sigma_d < X$ are considered as contaminants, with X=1 kpc for all the GCs at a distance $d<6$ kpc and X=5 kpc for GCs with $d>6$ kpc. In step (2) \texttt{PARSEC} isochrones \citep{Marigo2017} are used to define a region in the CMD where member stars are expected to be found. Only stars whose position in the CMD is compatible with the used isochrones are considered as possible members. We use $2\sigma$ as the maximum allowed distance from the star to the isochrone. Isochrones were chosen to have the known age, metallicity and reddening of each given cluster (from \citealp{Harris1996}, see also Table~\ref{tab:1}).

Member stars move coherently forming a clump in PM space. In step (3), we search the PM space for any overdensity of stars and determine which stars belong to it. We use \texttt{DBSCAN}, a density-based clustering algorithm, for this task \citep{Ester1996}. \texttt{DBSCAN} uses a distance, $\epsilon$, to determine whether two points are density-connected in the PM space. We define this distance as:

$\epsilon = \sigma_{1D}/2 + \left<  \sigma_{\nu} \right> + 3 \sigma(\sigma_{\nu}),$\\
where $\sigma_{1D}$ is the intrinsic velocity dispersion of the cluster in 1D converted to PM (we consider the central values of the velocity dispersion reported in \citealp{Baumgardt2018}, see also Table \ref{tab:1}), $\left< \sigma_{\nu} \right>$ is the average error in PM of the sample, and $\sigma(\sigma_{\nu})$ is the dispersion of the error in PM. Using the distance $\epsilon$, a cluster defined by \texttt{DBSCAN} must satisfy two properties: all points within the cluster must be mutually density-connected, and any point density-reachable from any point of the cluster is also part of the cluster. Using these criteria, \texttt{DBSCAN} identifies co-moving stars that are likely members of the cluster.

Each one of the steps from our pipeline produces a flag that is used to characterize the star membership probability from 0 to 3. We consider as cluster members those stars that simultaneously satisfy all of the selection criteria. In Table \ref{tab:1} we report the number of selected stars for each of the cleaned samples and in Figure \ref{fig:selection}, we show the cleaned data for NGC6656 and NGC6352, as an example.

The proper motion diagrams of Figure \ref{fig:selection} present spike features for high velocity stars.\footnote{Notice that these stars are excluded by our membership selection.} These spikes reflect low-quality astrometric measurements, specifically due to a dishomogeneous scanning pattern of Gaia, and can be identified through the astrometric quality parameter \texttt{variability\_periods\_used} (see e.g. \citealp{Lindegren2018}). The details of the effects on kinematics of these features are described in a follow up paper (Bianchini et al. in prep.). In this work, we verify that the rotation detections reported in the next sections do not depend on these features: kinematics subsamples that include a cut in the \texttt{variability\_periods\_used} give results fully consistent with the one obtained in the following sections.


\subsection{Global rotation measurements}
To measure the rotation on the plane-of-the-sky, we convert the stellar positions into cartesian coordinates centred on each cluster (centres from the Harris Catalogue, \citealp{Harris1996}), with the positive $x$-axis pointing toward West and the positive $y$-axis pointing toward North, using equation (1) from \citet{vandeVen2006}. The proper motions in the RA and Dec components are converted accordingly into a cartesian coordinate system using the orthographic projection reported in equation (2) of \citet{Helmi2018}, with positive $\mu_x$ in the direction of West.\footnote{This transformation is strictly necessary for objects with large apparent size and at large distance from the equatorial plane.} Finally, we transform the proper motions in radial and tangential components ($\mu_r$,$\mu_t$) in polar coordinates, with positive $\mu_r$ indicating velocities away from the cluster centre and positive $\mu_t$ indicating counterclockwise movements. In the rest of this paper we will focus on the analysis of the tangential component of the proper motions and analyze the radial component in Section \ref{sec:rad}.

\begin{figure}
\centering
\includegraphics[width=.5\textwidth]{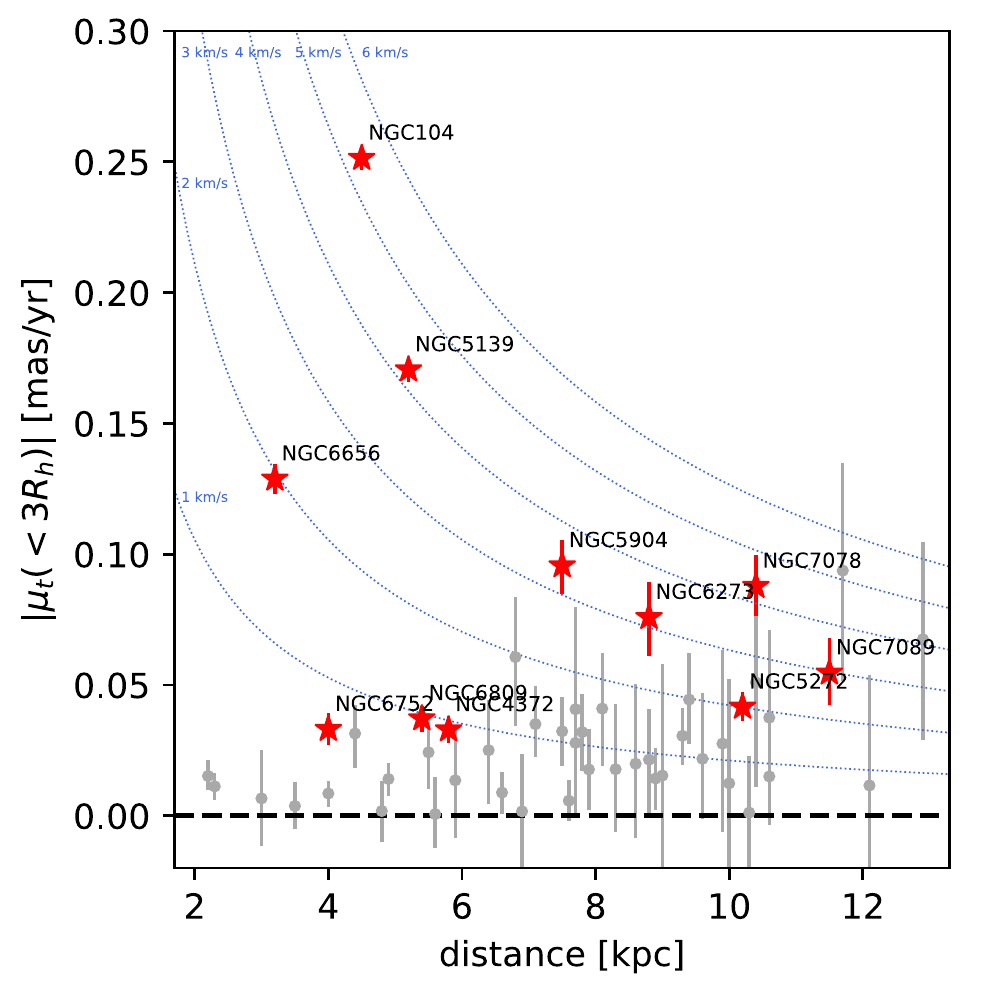}
\caption{Mean proper motion along the tangential component measured in a 3 half-light radii field-of-view around the cluster centres. The absolute value of $\mu_t(<3R_h)$ is plotted as a function of the distance of the clusters. Red stars indicate clusters for which rotation is detected at least with a 3-sigma confidence level. Dotted lines indicate  proper motion values in km s$^{-1}$ for a given distance ($v=4.74\,d\,\mu$, with $\mu$ measured in mas yr$^{-1}$, $v$ measured in km s$^{-1}$ and $d$ in kpc).}
\label{fig:3sigma}
\end{figure}

We compute the global rotation along the tangential component within $3R_h$ using equation \ref{eq:loglike}, as described in Section \ref{sec:2}. The results are reported in Table \ref{tab:1} and plotted in Figure \ref{fig:3sigma}. Out of the 51 GCs analyzed, 11 display an average value of $\mu_t$ significantly different than zero at a confidence  level $>3$-sigma. These clusters are indicated as red symbols in Figure \ref{fig:3sigma}.  We confirm the detection of rotation in NGC104, NGC5139, NGC5272, NGC5904, NGC6656, NGC6752, NGC6809, NGC7078 reported by \citet{Helmi2018}, and additionally we detect rotation in NGC4372, NGC6273 and NGC7089.

\citet{Helmi2018} and \citet{Lindegren2018} reported that systematics errors are present in \textit{Gaia} DR2, in particular a $\sim1^\circ$ small-scale component, related to the \textit{Gaia} scan pattern. Following the procedure outlined in Section 3.1 of \citet{vanderMarel2018} for M31, we ensure that these systematics do not induce an artificial rotation pattern in a typical GC field-of-view. We create a systematics pattern noise in a field-of-view of 10 arcmin and calculate the mean tangential proper motion. Our test suggests that within 10 arcmin the systematics are <0.5 $\mu$as~yr$^{-1}$, well below our typical errors on the mean tangential velocity ($\sim10$~$\mu$as yr$^{-1}$). Therefore we conclude that the small-scale systematic errors do not dominate our measurements.

The fraction of rotating GCs detected in line-of-sight studies is approximately $\sim50\%$ (\citealp{Kamann2018}). Here we detect rotation in a lower fraction of clusters, $\sim20\%$ of the sampled GCs. Assuming random inclinations for the rotation axes of GCs (inclination angle $i$ distributed as $\sin i\,di$), we expect to observe $\simeq70\%$ of clusters with inclinations higher than $45^\circ$ (with an average angle of $\simeq60^\circ$). This would imply that clusters are preferentially observed with an edge-on view, making a detection along the line-of-sight more likely, and in agreement with the higher fraction of rotating clusters reported in \citet{Kamann2018}. However, we note that the study of \citet{Kamann2018} and the one in this work refer to different parts of the clusters, with the \citet{Kamann2018} sample being more centrally concentrated. Additionally, the detection limits of Gaia DR2 could likely limit the ability of measuring low-rotation signals for clusters at larger distances. This could further explain the different fractions of rotating clusters detected in the two works.

In addition to the clusters for which we detected a signal of rotation at a $>3$-sigma level, we also report detection of rotation at a 2-sigma confidence level for 11 GCs: NGC3201, NGC5286, NGC6121, NGC6205, NGC6254, NGC6266, NGC6397, NGC6402, NGC6539, NGC6541, NGC6779. We identify these clusters as the best targets for follow-up plane-of-the-sky rotation studies when higher quality data will be available (e.g. the following \textit{Gaia} data release).

\vspace{1cm}

\begin{figure*}
\centering
\includegraphics[width=.92\textwidth]{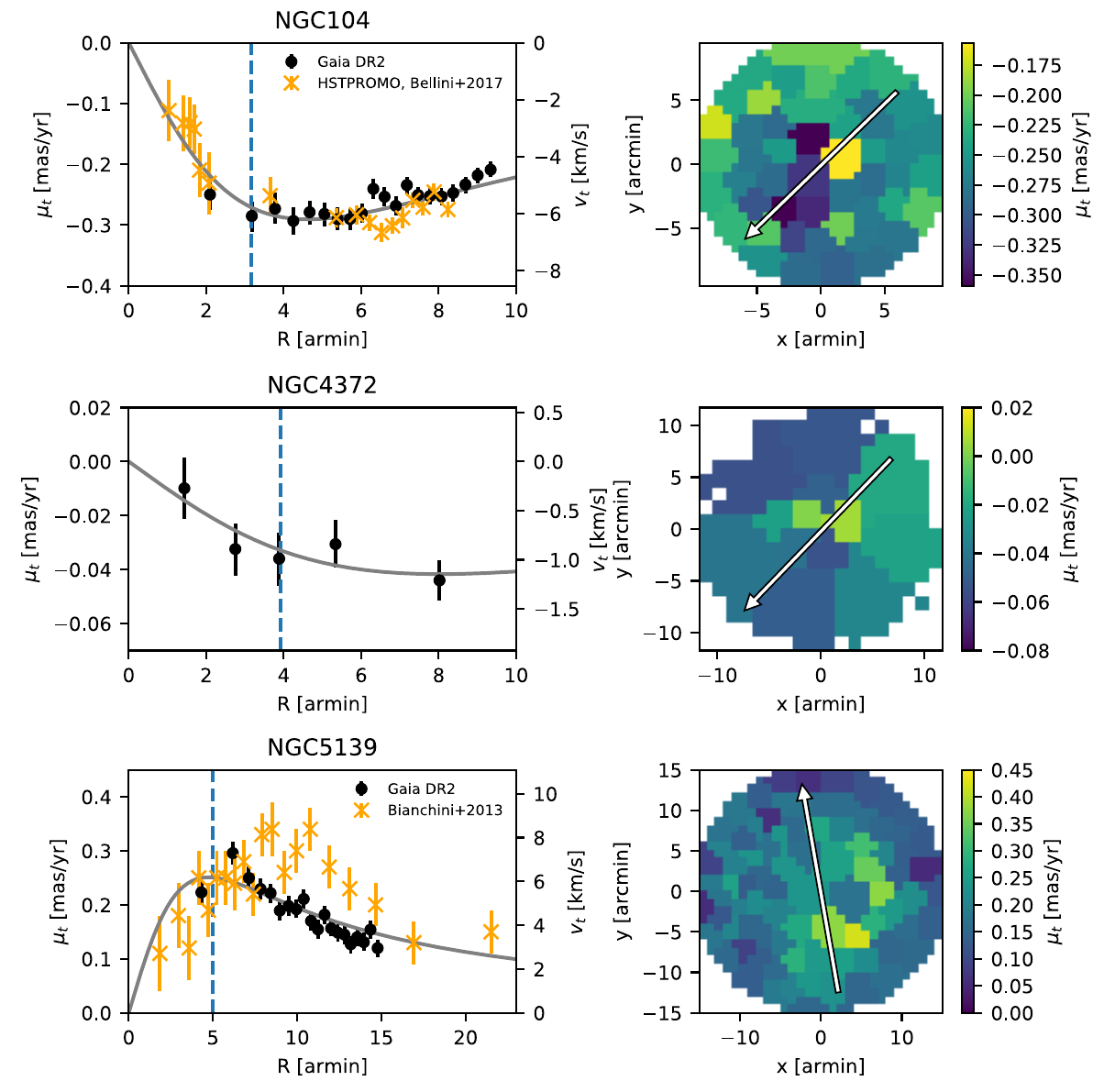}
\caption{Rotation curves and rotation maps for the 11 clusters for which a global value of rotation was detected (see also following Figures \ref{fig:rot_curve2} and \ref{fig:rot_curve3}.). Black points refer to the results from this work, while orange symbols are rotation profiles from the literature. The grey curves correspond to the best fit rotation profile (equation \ref{eq:rotation_profile}) while the dotted blue vertical lines indicate the position of the half-light radius $R_h$. \textit{Positive values in the rotation maps indicate a counterclockwise rotation on the plane-of-the-sky.} The arrows indicate the position angle of the rotation axis derived from line-of-sight studies (see Table \ref{tab:inclination}).}
\label{fig:rot_curve1}
\end{figure*}
\begin{figure*}
\centering
\includegraphics[width=0.92\textwidth]{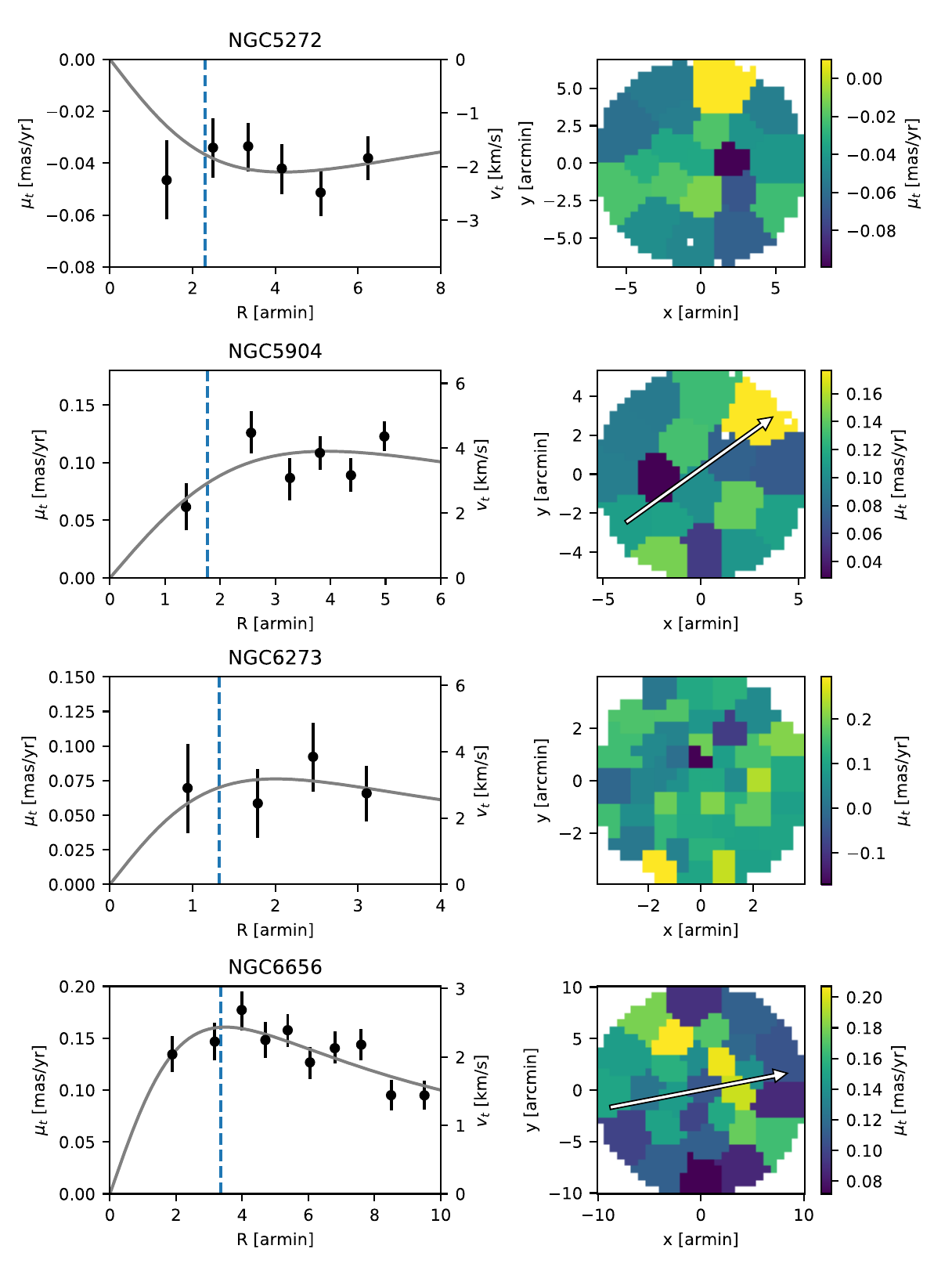}
\caption{As in Figure \ref{fig:rot_curve1}.}
\label{fig:rot_curve2}
\end{figure*}
\begin{figure*}
\centering
\includegraphics[width=.92\textwidth]{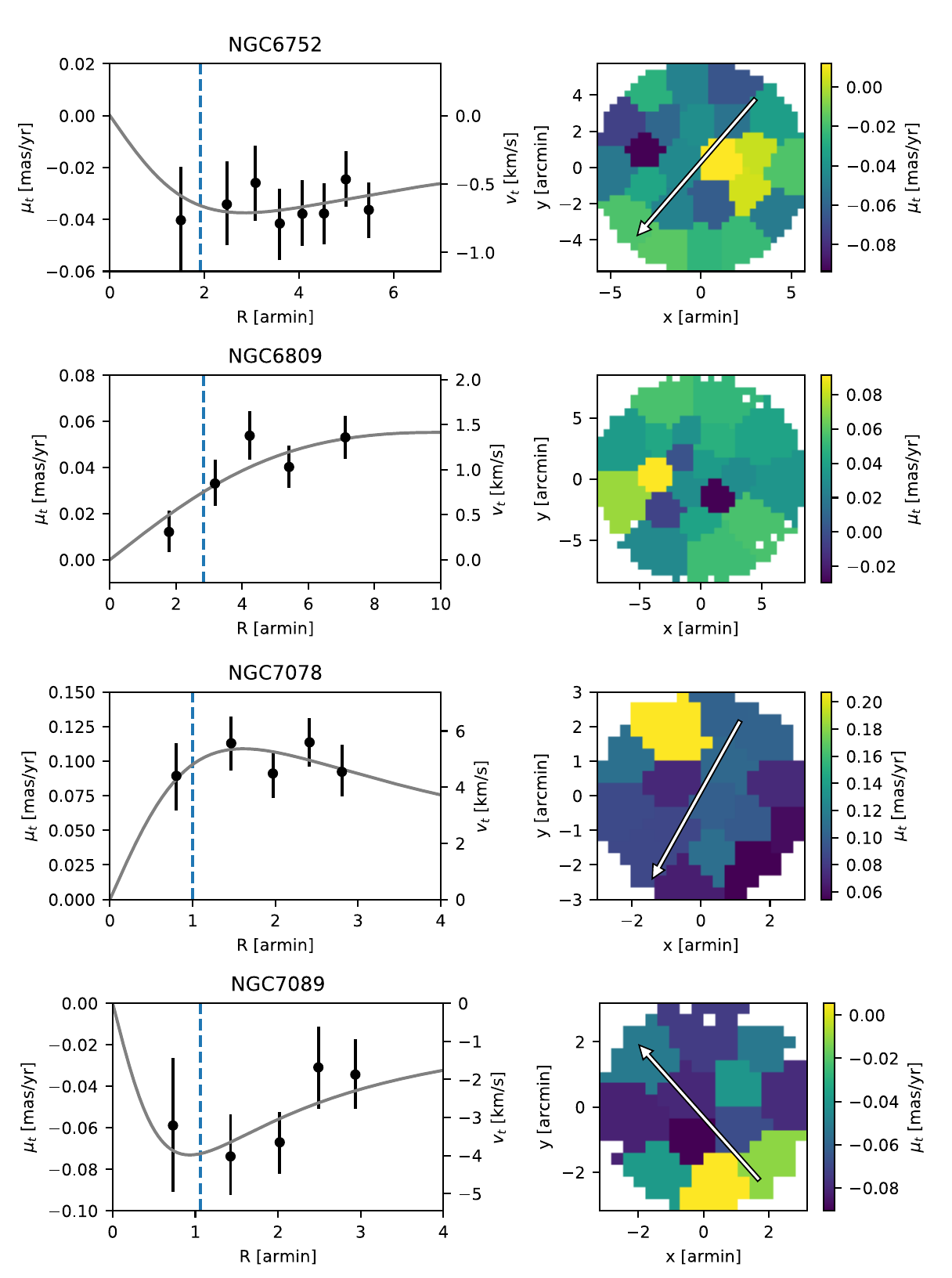}
\caption{As in Figure \ref{fig:rot_curve1}.}
\label{fig:rot_curve3}
\end{figure*}
 
\subsection{Rotation maps and profiles}
\label{sec:rot_maps}
For the 11 clusters for which we reported a global detection of rotation ($>3$-sigma detection), we perform a more detailed analysis in order to extract the spatial distribution of the rotation signal; in particular, we construct proper motion rotation profiles and two-dimensional rotation maps.

Here we note that the procedure to select member stars followed in Section \ref{sec:selection} does not take into consideration the radial-dependency of the velocity dispersion $\sigma$. For this reason, we implement a further radial-dependent $\sigma$-clipping to ensure that the contamination from background and foreground stars is successfully minimized, especially in the outer areas of the fields-of-view where the relative density of contaminants is higher.
For each cluster, we divide the datasets into 5 radial bins containing the same number of stars, we calculate the velocity dispersion in the tangential component for each radial bin (using equations \ref{eq:loglike} and \ref{likelihood}) and exclude the stars with tangential velocities greater than $3\sigma$. 

The newly cleaned datasets are used to calculate the rotation profiles, computed by dividing the field-of-view into concentric radial bins containing the same number of stars. For each bin we use equations \ref{eq:loglike} and \ref{likelihood} to calculate the mean tangential velocity and associated errors. The details of the profiles of each cluster are reported in Tables \ref{tab:profiles_1} and \ref{tab:profiles_2}.

The two-dimensional velocity maps are computed by first dividing the fields-of-view in a grid of equally sized bins (grid of 30$\times$30 and 50$\times$50 bins for those clusters with low and high number of stars, respectively) and then by using the Voronoi binning algorithm (\citealp{CappellariCopin2003}) that re-bins the field-of-view into bins with an equal number of stars (with at least 100 stars per bin). For each bin we calculate the mean velocity along the tangential component and the associated errors, as described in Section \ref{sec:2} above. The rotation velocity profiles and maps are reported in Figures \ref{fig:rot_curve1}, \ref{fig:rot_curve2} and~\ref{fig:rot_curve3}.

\section{Results}
Figures \ref{fig:rot_curve1}-\ref{fig:rot_curve3} show the typical differential rotation pattern characterized by lower values closer to the centre, a peak in the intermediate region and a decrease toward the outer part. For 6 out of the 11 rotating GCs we find counterclockwise rotation (positive values of the mean $\mu_t$) while the other 5 display clockwise rotation. This is consistent with a random distribution of rotation patterns, indicating that the observed rotation is not due to systematics. To quantify the contribution of rotation, we fit the profiles using the empirical profile provided by equation~\ref{eq:rotation_profile}. With the fits we are able to quantify the position and the value of the rotation peak, as reported in Table~\ref{tab:1}. The detected maximum rotation signal ranges between  $\sim0.5$ km s$^{-1}$ for NGC4372 and $\sim6$ km s$^{-1}$ for NGC104 and NGC5139, and the rotation peaks are located within 0.5-2 $R_h$ from the cluster centres (consistent with the findings in line-of-sight studies, e.g. \citealp{Bianchini2013,Ferraro2018}; see also \citealp{Tiongco2017} for the expectation from $N$-body modelling). Note that a proper crowding treatment was not implemented in \textit{Gaia} DR2 causing the lack of observed stars in very crowded regions. This affects the central regions of most of our clusters, were no data is available at small radii ($<1$ arcmin).

We perform a comparison of the results obtained in this work with the previous proper motion studies of NGC104 and NGC5139.
For NGC104, we compare our results with the rotation profile of \citet{Bellini2017} (HSTPROMO collaboration) measured with \textit{HST} proper motions using the background stars of the Small Magellanic Cloud as an absolute reference frame. The first panel of Fig. \ref{fig:rot_curve1} demonstrates the excellent agreement between the two datasets, indicating the optimal capability of \textit{Gaia} to detect rotation signatures in proper motions.

For NGC5139, we compare our results with the proper motion rotation profile used in \citet{Bianchini2013}, indicating a good agreement between the two datasets (see third panel of Fig. \ref{fig:rot_curve1}). The discrepancies visible around 10-15 arcmin can be ascribed to the construction procedure of the literature sample (originally described in \citealp{vanLeeuwen2000,LePoole2002}). In fact, these ground-based proper motions do not directly measure the internal rotation of NGC5139, since the astrometric reduction process removes any signature of solid body rotation in the cluster. A correction to the dataset has been applied to recover the internal rotation signal, as described in Section 4 of \citet{vandeVen2006} exploiting a general relation for axisymmetric objects.


\begin{table}
\tabcolsep=0.10cm
\begin{center}
\caption{Rotation profiles derived in this paper and used in Figure \ref{fig:rot_curve1}, \ref{fig:rot_curve2} and \ref{fig:rot_curve3}.}
\begin{tabular}{llll}
\hline\hline
R& $\mu_t$ & $e_{low}$ & $e_{up}$ \\
arcmin& mas yr$^{-1}$ & mas yr$^{-1}$ & mas yr$^{-1}$\\
\hline
NGC104 & $N_{bin}=1355$ & & \\
\hline
      2.10   &  -0.250   &   0.026   &   0.026\\
      3.18   &  -0.285   &   0.027   &   0.024\\
      3.78   &  -0.273   &   0.024   &   0.026\\
      4.25   &  -0.293   &   0.023   &   0.023\\
      4.67   &  -0.279   &   0.021   &   0.018\\
      5.05   &  -0.282   &   0.020   &   0.019\\
      5.39   &  -0.290   &   0.018   &   0.019\\
      5.71   &  -0.290   &   0.018   &   0.018\\
      6.01   &  -0.281   &   0.016   &   0.016\\
      6.31   &  -0.241   &   0.016   &   0.016\\
      6.60   &  -0.254   &   0.016   &   0.016\\
      6.89   &  -0.269   &   0.015   &   0.016\\
      7.18   &  -0.234   &   0.015   &   0.014\\
      7.47   &  -0.251   &   0.014   &   0.014\\
      7.77   &  -0.253   &   0.014   &   0.014\\
      8.07   &  -0.253   &   0.014   &   0.014\\
      8.38   &  -0.247   &   0.014   &   0.014\\
      8.69   &  -0.234   &   0.014   &   0.013\\
      9.01   &  -0.218   &   0.014   &   0.012\\
      9.34   &  -0.209  &    0.013   &   0.013\\

\hline
NGC4372 & $N_{bin}=1097$ & & \\
\hline
      1.44  &   -0.010   &   0.011   &   0.011\\
      2.75  &   -0.032    &  0.010   &   0.009\\
      3.88  &   -0.036   &   0.010   &   0.010\\
      5.34   &  -0.031    &  0.009   &   0.009\\
      8.02   &  -0.044    &  0.007   &   0.007\\

\hline
NGC5139 & $N_{bin}=1779$ & & \\
\hline
      4.31  &    0.223   &   0.021   &   0.021\\
      6.18   &   0.296   &   0.021   &   0.022\\
      7.13   &   0.250   &   0.020   &   0.021\\
      7.82   &   0.229   &   0.019    &  0.019\\
      8.42   &   0.222   &   0.018   &   0.018\\
      8.97   &   0.189   &   0.018   &   0.017\\
      9.49   &   0.198   &   0.018   &   0.018\\
      9.95   &   0.192    &  0.017   &   0.018\\
     10.39  &    0.211   &   0.017  &    0.017\\
     10.82  &    0.170   &   0.018   &   0.017\\
     11.23  &    0.155   &   0.017   &   0.018\\
     11.62   &   0.182   &   0.017   &   0.016\\
     12.02   &   0.157   &   0.015  &    0.016\\
     12.42   &   0.149   &   0.017  &    0.018\\
     12.80   &   0.145   &   0.017  &    0.017\\
     13.18   &   0.127   &   0.017   &   0.015\\
     13.56   &   0.140   &   0.016   &   0.017\\
     13.95  &    0.131   &   0.016   &   0.017\\
     14.35  &    0.154   &   0.017   &   0.017\\
     14.77   &   0.120   &   0.016   &   0.016\\

 \hline
NGC5272 & $N_{bin}=1079$ & & \\
\hline
      1.37  &   -0.047   &   0.015   &   0.015\\
      2.49  &   -0.034   &   0.012   &   0.011\\
      3.34  &   -0.034   &   0.010   &   0.009\\
      4.16  &   -0.042   &   0.010   &   0.009\\
      5.10  &   -0.051   &   0.009   &   0.009\\
      6.24  &   -0.038   &   0.008   &   0.008\\
   \hline
\end{tabular}
\label{tab:profiles_1}
\end{center}
\end{table}

\begin{table}
\tabcolsep=0.10cm
\begin{center}
\caption{Rotation profiles derived in this paper and used in Figure \ref{fig:rot_curve1}, \ref{fig:rot_curve2} and \ref{fig:rot_curve3}.}
\begin{tabular}{llll}
\hline\hline
R& $\mu_t$ & $e_{low}$ & $e_{up}$ \\
arcmin& mas yr$^{-1}$ & mas yr$^{-1}$ & mas yr$^{-1}$\\

\hline
NGC5904 & $N_{bin}=996$ & & \\
\hline
      1.38   &   0.062   &   0.020   &   0.021\\
      2.56   &   0.126   &   0.018   &   0.019\\
      3.27   &   0.087   &   0.020   &   0.018\\
      3.81   &   0.108   &   0.015   &   0.015\\
      4.37   &   0.089   &   0.015   &   0.015\\
      4.98   &   0.123   &   0.013   &   0.013\\

\hline
NGC6273 & $N_{bin}=491$ & & \\
\hline
      0.94   &   0.070   &   0.032   &   0.032 \\
      1.79  &    0.059   &   0.025   &   0.025 \\
      2.46  &    0.092   &   0.025    &  0.024 \\
      3.10  &    0.066   &   0.020   &   0.020 \\

\hline
NGC6656 & $N_{bin}=1181$ & & \\
\hline
      1.89   &   0.134   &   0.017   &   0.018\\
      3.17   &   0.147   &   0.018   &   0.018\\
      3.98   &   0.177   &   0.020   &   0.018\\
      4.70   &   0.148   &   0.017   &   0.017\\
      5.37   &   0.158   &   0.016   &   0.015\\
      6.05   &   0.127   &   0.016   &   0.015\\
      6.80   &   0.140   &   0.015   &   0.016\\
      7.60   &   0.144   &   0.015   &   0.015\\
      8.50   &   0.095   &   0.015   &   0.014\\
      9.51   &   0.095   &   0.014   &   0.014\\

\hline
NGC6752 & $N_{bin}=1291$ & & \\
\hline
      1.50    & -0.040   &   0.020   &   0.021\\
      2.47    & -0.034   &   0.016   &   0.017\\
      3.08    & -0.026   &   0.015   &   0.014\\
      3.59    & -0.042   &   0.014   &   0.013\\
      4.07    & -0.038    &  0.012   &   0.013\\
      4.53    & -0.038   &   0.012   &   0.012\\
      4.99    & -0.025   &   0.010   &   0.011\\
      5.48    & -0.036   &   0.011   &   0.011\\

\hline
NGC6809 & $N_{bin}=1302$ & & \\
\hline
      1.79    &  0.012   &   0.009    &  0.009\\
      3.18    &  0.033   &   0.010    &  0.010\\
      4.23    &  0.054   &   0.011    &  0.011\\
      5.42    &  0.040   &   0.009    &  0.009\\
      7.12    &  0.053   &   0.010    &  0.009\\

\hline
NGC7078 & $N_{bin}=342$ & & \\
\hline
      0.80   &   0.089    &  0.025   &   0.024\\
      1.47   &   0.113    &  0.020    &  0.019\\
      1.97   &   0.091   &   0.018   &   0.016\\
      2.41   &   0.114    &  0.017    &  0.018\\
      2.81   &   0.092   &   0.018   &   0.019\\

\hline
NGC7089 & $N_{bin}=276$ & & \\
\hline
      0.73   &  -0.059     & 0.032   &   0.033\\
      1.43   &  -0.074    &  0.019   &   0.020\\
      2.02   &  -0.067    &  0.015   &   0.015\\
      2.49   &  -0.031    &  0.020   &   0.020\\
      2.93   &  -0.034   &  0.017    &  0.017\\
\hline

\end{tabular}
\label{tab:profiles_2}
\end{center}
\end{table}

\subsection{Comparison with \citet{Helmi2018}}
\begin{figure}
\centering
\includegraphics[width=.46\textwidth]{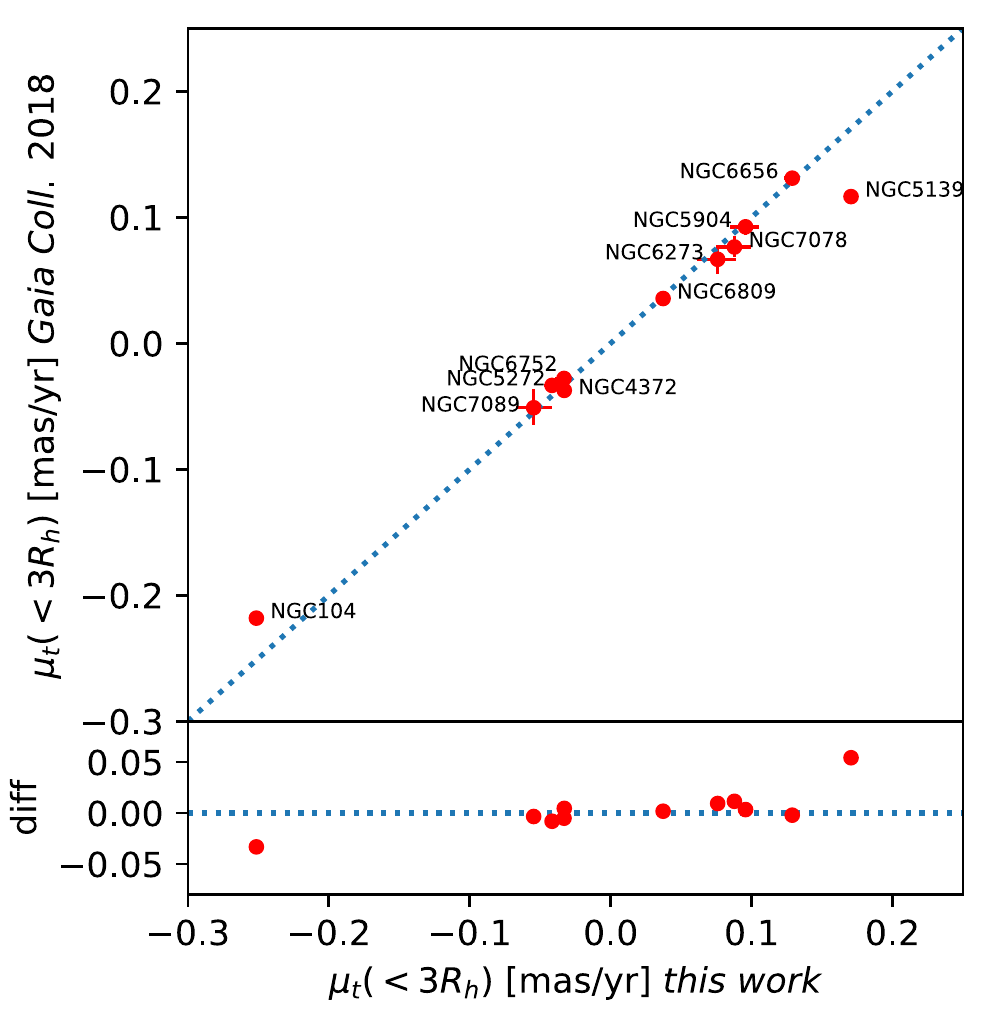}
\caption{Comparison between the global values of rotation within $3R_h$ from our work and from the stars in common with \citet{Helmi2018} samples. The bottom panel shows the difference between the two datasets, in mas yr$^{-1}$.}
\label{fig:comparison}
\end{figure}

In addition to the comparisons with previous proper motion rotation studies presented above, we perform a comparison with the results presented in \citet{Helmi2018}, as part of the \textit{Gaia} DR2. In particular, for each of the 11 clusters for which we detected rotation and constructed rotation profiles, we assure that our membership selection procedure (Section \ref{sec:selection}) gives results consistent with the membership selection used in \citet{Helmi2018}. Our sample is restricted to the central $3R_h$ area around the cluster centres, while the clean catalogs of \citet{Helmi2018} extend to the tidal radii of the clusters; moreover, our selection procedure takes into consideration the radial dependency of the velocity dispersion (see Section \ref{sec:rot_maps}). We select the stars in common between the two clean catalogs (using the members in \citealp{Helmi2018} reported in their Table D.3), and apply our analysis to construct rotation maps and rotation profiles. The rotation profiles and maps obtained are consistent with the one presented in this paper, stressing the robustness of our results. In Figure \ref{fig:comparison}, we show the comparison between the global rotation values within $3R_h$ for our work and for the stars in common with the \citet{Helmi2018} catalog. A clear outlier is NGC5139, due to the lack of coverage in the central region of the \citet{Helmi2018} selected data, that do not sample the area around the rotation peak (see their Figure A.6).

\begin{figure*}
\centering
\includegraphics[width=.92\textwidth]{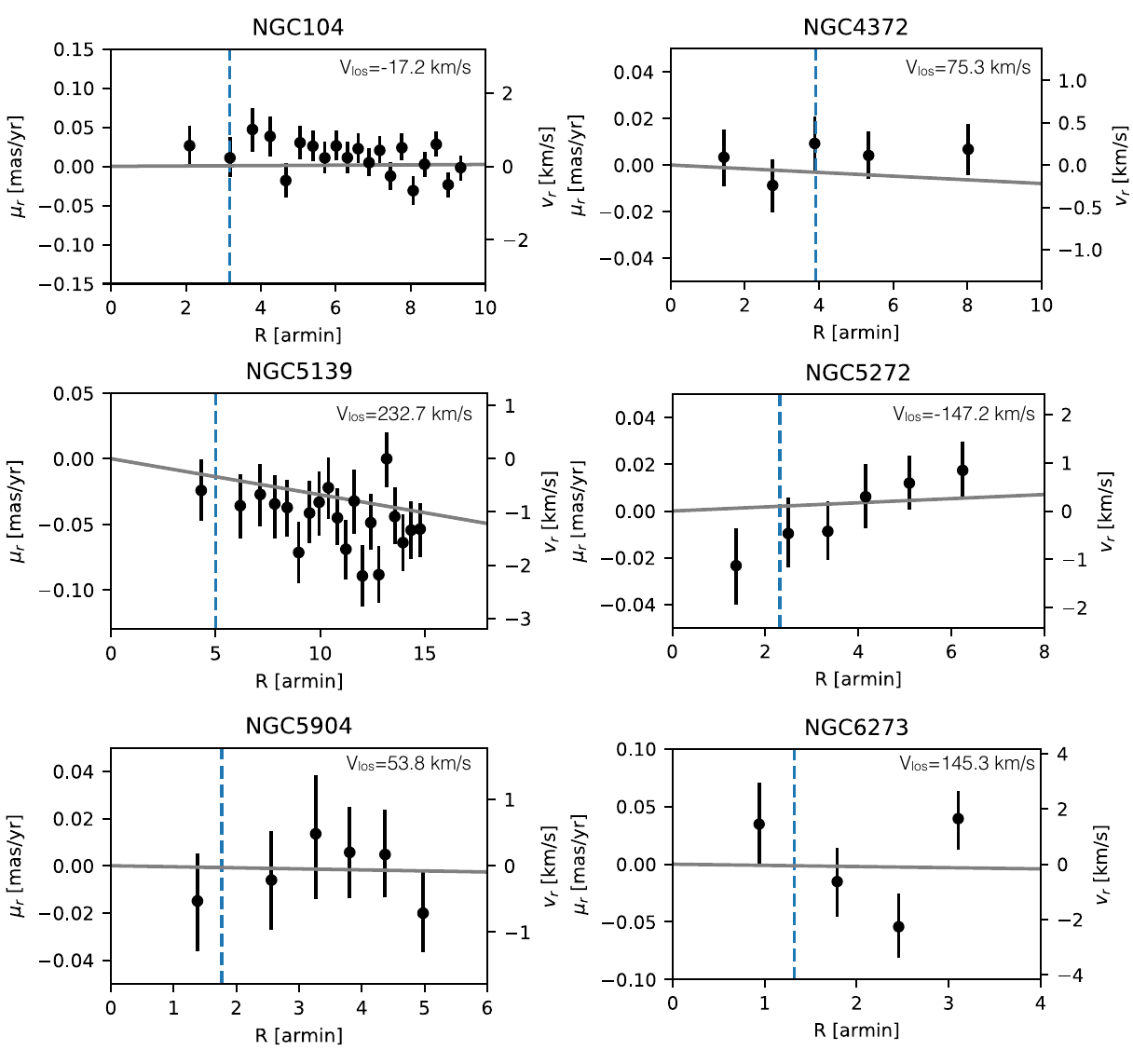}
\caption{Profiles of average $\mu_r$ for the 11 clusters for which rotation has been detected. The vertical dashed lines indicate the half-light radii, while the solid grey lines indicate the expected perspective expansion/contraction given by equation \ref{eq:expansion}, calculated using the values of the line-of-sight systemic motion of a GC specified on the top right of each plot (from \citealp{Baumgardt2018}).}
\label{fig:vr_curve}
\end{figure*}
\begin{figure*}
\centering
\includegraphics[width=.92\textwidth]{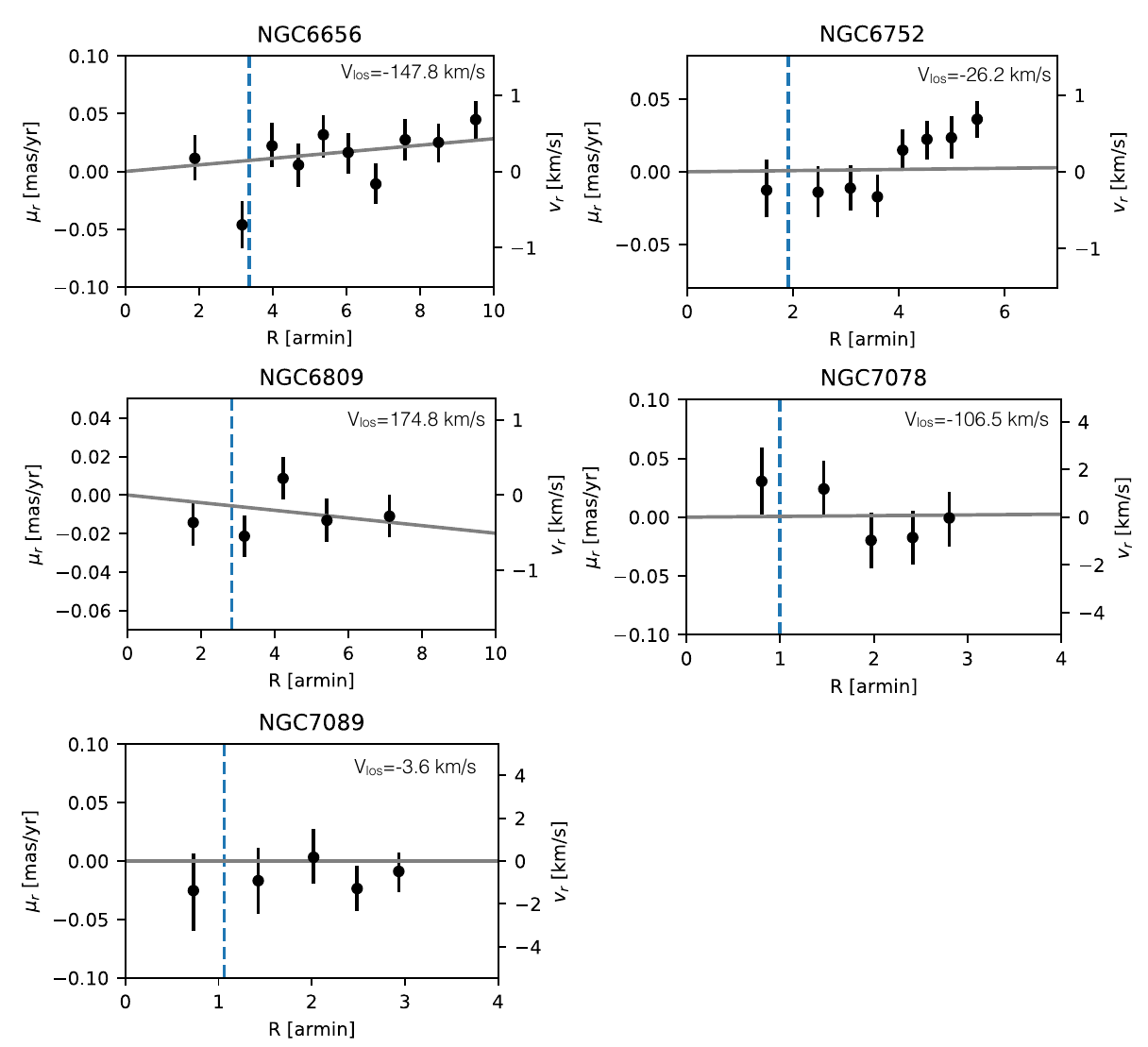}
\caption{As in Figure \ref{fig:vr_curve}.}
\label{fig:vr_curve2}
\end{figure*}

\subsection{Mean proper motion along the radial component}
\label{sec:rad}
So far we have concentrated exclusively on the tangential component of the proper motions $\mu_t$. For a consistency check, we compute the profile of the radial component of the proper motions $\mu_r$ using the same procedure described in Section \ref{sec:rot_maps}. Values of $\mu_r$ significantly different than zero would indicate a contraction/expansion of the cluster (positive values of $\mu_r$ indicate expansion). A systematic contraction/expansion is expected due to the bulk motion of the GC along the line-of-sight, as given by equation 6 of \citet{vandeVen2006}
\begin{equation}
\mu_r=-6.1363\times10^{-5}\, V_{los}\,R/d \quad \mathrm{mas\,yr^{-1}},
\label{eq:expansion}
\end{equation}
where $V_{los}$ is the systemic velocity of the cluster along the line-of-sight in km s$^{-1}$, $R$ is the distance to the cluster centre in arcmin and $d$ is the distance of the cluster to the sun in kpc.

In Figures \ref{fig:vr_curve} and \ref{fig:vr_curve2} we show the profiles of the mean $\mu_r$ and overplot the expected perspective contraction/expansion from equation \ref{eq:expansion}, using as values of $V_{los}$ the one reported in \citet{Baumgardt2018}. Despite the scatter of the data due to small mean values of the order of $<1$ km s$^{-1}$, our results show consistent trends with the expected contraction/expansion. This results indicates that if data systematics are present they are below the random uncertainties of the data. A detailed analysis of the impact of \textit{Gaia} DR2 error systematics on the mean radial proper motion component profiles will be presented in an upcoming paper (Bianchini et al. in prep.).

\subsection{V/$\sigma$ parameter}
\begin{figure*}
\centering
\includegraphics[width=.95\textwidth]{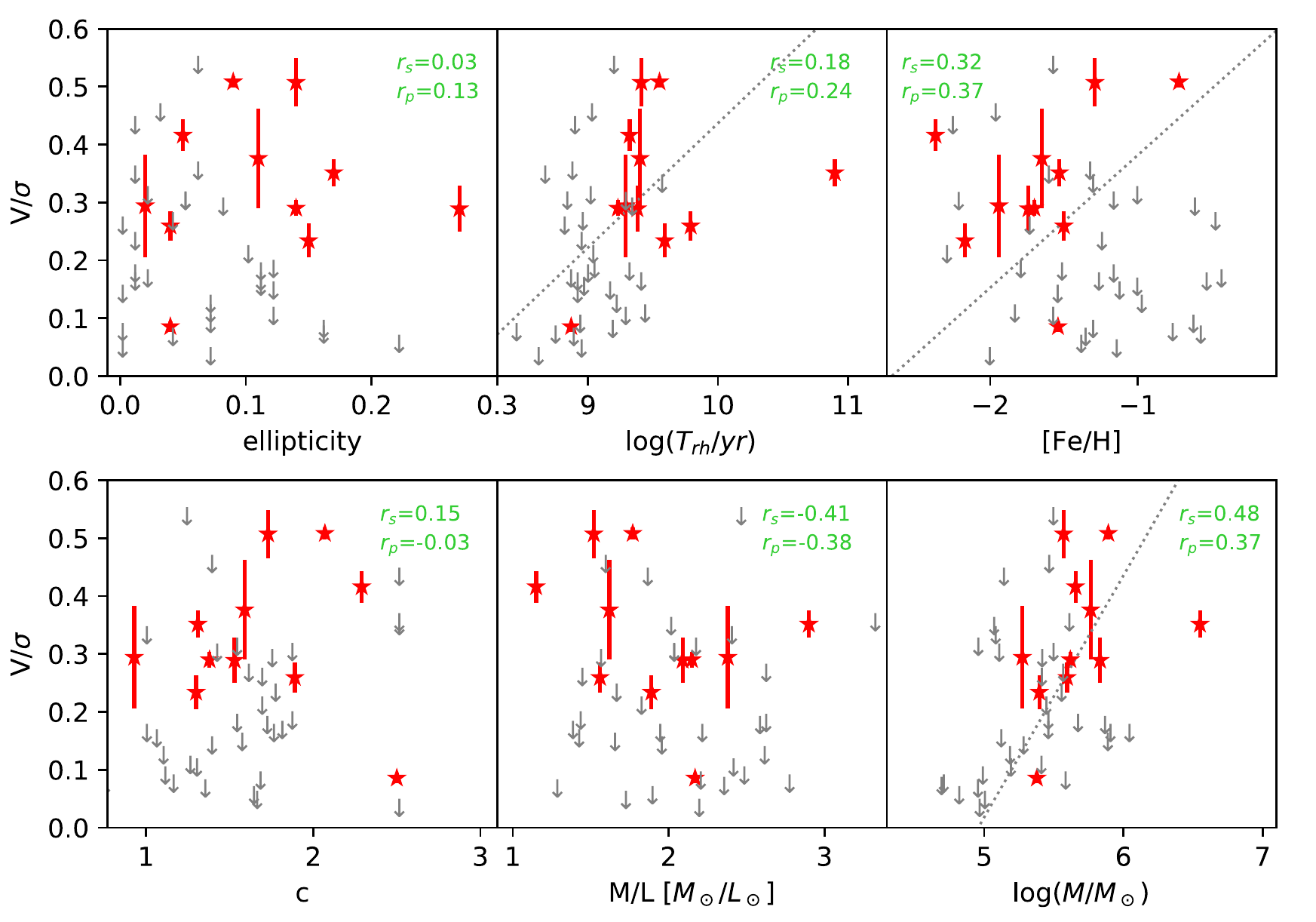}
\caption{$V/\sigma$ parameter as a function of the clusters ellipticity, half-mass relaxation time $T_{rh}$, metallicity [Fe/H], concentration c, mass-to-light ratio M/L and total cluster mass M. The red symbols indicate the 11 GCs for which rotation has been detected at a >3-sigma level, while grey symbols are the upper limits derived for all the other clusters. At the top of each plot we indicate the Spearman's rank correlation coefficient $r_s$ and the Pearson correlation coefficient $r_p$ and we overplot a linear fit to the data when an indication of correlation is found. The correlation tests and the fits are performed only on the clusters with >3-sigma rotation detection.}
\label{fig:vsigma}
\end{figure*}

A simple tool to assess the importance of internal rotation in shaping stellar systems is the study of the $V/\sigma$ parameter (e.g., \citealp{Davies1983,Emsellem2011,Bianchini2013}), defined as the ratio between the peak of the rotation (from this work) and the central velocity dispersion (from line-of-sight measurements, \citealp{Baumgardt2018}).\footnote{Note that an alternative method to define the $V/\sigma$ parameter has been presented in \citet{Emsellem2011} following the definition provided in \citet{Binney2005}.} The $V/\sigma$ values measured in this work are reported in Table \ref{tab:1} and display values between 0.08 for NGC6752 and $\sim$0.5 for NGC104 and NGC5904. These values provide the first measurements of $V/\sigma$ on the plane-of-the sky for a large sample of clusters.

For the clusters for which a global rotation signal was not detected with a 3-sigma confidence level, we estimate an upper limit of the $V/\sigma$ as the upper 1-sigma limit of the mean tangential proper motion measured within 3 half-light radii, divided by the central velocity dispersion (as reported in Table \ref{tab:1}).
In Figure \ref{fig:vsigma} we relate the $V/\sigma$ parameters derived in this work with a number of GC properties, namely the cluster ellipticity, the half-mass relaxation time $T_{rh}$, metallicity [Fe/H], concentration c (from \citealp{Harris1996}), the mass-to-light ratio M/L and total cluster mass M (from \citealp{Baumgardt2018}). For each plot we perform a Spearman rank test and calculate the Pearson correlation coefficient indicating the correlation between $V/\sigma$ and the different cluster properties (we indicate the two correlation coefficients with $r_s$ and $r_p$, respectively). The correlation tests are performed only for the objects with a 3-sigma rotation detection (red symbols in Figure \ref{fig:vsigma}). We find weak to moderate correlations, the strongest of which are between the $V/\sigma$ and the metallicity, the relaxation time, and the total mass.   

The correlation between line-of-sight rotation and metallicity was reported in \citet{Bellazzini2012} and \citet{Lardo2015}, indicating higher rotation for metal-rich clusters ($r_s=0.423$ and $r_s=0.32$ for \citet{Lardo2015} and this work, respectively). As pointed out by \citet{Bellazzini2012} this correlation, if confirmed, could be be indicative of the earliest phases of GC formation, where higher metallicity would imply higher efficiency in energy dissipation (see \citealp{Bekki2010,Bekki2011}). This correlation was not confirmed in \citet{Kimmig2015} and neither in \citet{Kamann2018}, whose data were however limited to the central areas only.  When we include all the clusters (grey upper limits in Figure \ref{fig:vsigma}) the hint of correlation between $V/\sigma$ and metallicity is not present anymore. We argue that a study of the intrinsic three-dimensional rotation signature is necessary to reach a robust assessment of this correlation, to avoid the complications derived by projection effects. 

On the other hand, the correlation between $V/\sigma$ and the relaxation time is particularly interesting since it connects the current properties of a cluster with its long-term dynamical evolution. Angular momentum is expected to be transported outwards in a stellar systems while it dynamically relaxes. This naturally causes the decline of the rotation peak with time (see \citealp{Tiongco2017} and references therein). Therefore, clusters with longer relaxation times are expected to retain more efficiently their primordial signatures of internal rotation. This result confirms the finding reported in \citet{Kamann2018} ($r_s=0.55$ and $r_s=0.18$ for \citet{Kamann2018} and this work, respectively) and highlights the importance of internal rotation in the earliest phases of GC formation, despite the moderate amplitude of rotation measured in present-day GCs. We notice that the correlation between the $V/\sigma$ and the total mass reinforces the validity of the discussion above, since clusters with higher mass display longer relaxation times. When we include all the clusters (grey upper limits in Figure \ref{fig:vsigma}) the correlations are strengthened, given further support to our previous findings.


\subsection{Estimate of the inclination of the rotation axis}
\label{sec:inc}

\begin{table*}
\begin{center}
\caption{Estimate of the inclination of the rotation axis derived from a comparison between tangential PMs along the major and minor axis (equation \ref{eq:i_maps}) and from a comparison between PM rotation and line-of-sight rotation (equation \ref{eq:i_vlos}). $\phi$ represents the position angle of the rotation axis (measured from North to East) from line-of-sight studies ($^{(a)}$ \citealp{Kamann2018}, $^{(b)}$ \citealp{Kacharov2014}), $\Delta$ is the width of the slit along the major and minor axis,  $\mu^{min}_t$ and $\mu^{maj}_t$ are the mean values of the tangential proper motions along the minor/major axis.}
\begin{tabular}{lcccccc}
\hline\hline
& $\phi$ & $\Delta$&$\mu_t^{min}$ & $\mu_t^{maj}$& $i$ (eq. \ref{eq:i_maps}) & $i$ (eq. \ref{eq:i_vlos})\\
& deg & arcmin&mas yr$^{-1}$ & mas yr$^{-1}$ & deg &deg\\

\hline

NGC104 & 134.1 $^{(a)}$ & 1.0 &$-0.27\pm{0.01}$ & $-0.27\pm{0.01}$&<45 & $26\pm1$\\
NGC4372 &136 $^{(b)}$ & 1.5 & $-0.025\pm{0.010}$&$-0.013\pm{0.010}$ &$\simeq60$ & $49\pm2$\\
NGC5139 & 9.9 $^{(a)}$ & 0.5 &$0.220\pm{0.024}$ & $0.163\pm{0.022}$&$\simeq40$& $44\pm2$ \\
NGC5904 & -54.3 $^{(a)}$ & 0.5& $0.094\pm{0.022}$& $0.090\pm{0.021}$&<45 & $38\pm3$\\
NGC6656 & -79.1 $^{(a)}$ & 0.5 &$0.125\pm{0.021}$ &$0.104\pm{0.020}$ &$\simeq35$ &\\
NGC6752 & 139.1 $^{(a)}$ & 0.5 &$-0.050\pm{0.018}$ &$-0.028\pm{0.016}$ &$\simeq55$ &\\
NGC7078 &150.9 $^{(a)}$ &0.25 &$0.105\pm{0.023}$ & $0.110\pm{0.024}$&<45 & $29\pm3$\\
NGC7089 & 41.7 $^{(a)}$ &0.3 & $-0.073\pm{0.039}$&$-0.020\pm{0.028}$ & $\simeq75$ &\\

\hline
\end{tabular}
\label{tab:inclination}
\end{center}
\end{table*}

In this section we provide a first estimate of the inclination of the rotation axis with respect the line-of-sight for a subset of the GCs presented in this work. We apply two simple methods to assess the inclination $i$: (1) we compare the rotation on the plane-of-the-sky to the rotation along the line-of-sight; (2) we exploit the two-dimensional rotation maps presented in Section \ref{sec:rot_maps}.
 
In general, line-of-sight values of $V/\sigma$ higher than the one measured on the plane-of-the-sky imply that a cluster is viewed close to an edge-on projection, that is the rotation axis is almost perpendicular to the line-of-sight ($i>45^\circ$). Vice versa, clusters with a rotation axis pointing close to the direction of the line-of-sight ($i<45^\circ$) will display higher values of $V/\sigma$ on the plane-of-the-sky. This comparison is straightforward only for a limited number of clusters, since only global values of the rotation strength are usually reported and the peak of the rotation curve is not characterized (e.g. \citealp{Kamann2018}). 

NGC104, NGC5904, and NGC7078 have $V/\sigma$ on the plane-of-the-sky measured in this work of 0.51, 0.51 and 0.42, respectively, and of 0.25 (\citealp{Bianchini2013}), 0.4 (\citealp{Lanzoni2018}), and 0.23 (\citealp{Bianchini2013}) along the line-of-sight. This indicates that the rotation axis is significantly inclined and the intrinsic contribution of internal rotation is higher than the one so far inferred from line-of-sight data alone. This was already reported in \citet{Bellini2017} for NGC104, where detailed axisymmetric rotating models were applied to a three-dimensional kinematic dataset leading to an estimate of the inclination of the rotation axis of $i\sim30$. This result was not previously known for NGC7078 and NGC5904 (\citealp{Lanzoni2018}).
For NGC4372 and  NGC5139 the line-of-sight $V/\sigma$ (0.26, \citealp{Kacharov2014}, 0.34, \citealp{Bianchini2013}) are comparable to the one on the plane-of-the-sky (0.23 and 0.35, respectively), indicating that the systems are likely viewed on an angle $i\sim45^\circ$ (see \citealp{vandeVen2006} for NGC5139).

A more quantitative analysis can be carried out using the following general relation for axisymmetric systems (see equation 8 from \citealp{vandeVen2006})
\begin{equation}
\langle v_{los}\rangle(x',y')=4.74\, d \tan i \,\langle\mu_{y^{'}}\rangle(x',y')
\label{eq:i_vlos}
\end{equation}
connecting locally the average line-of-sight velocity $v_{los}$ to the average proper motion component along the rotation axis of the system $\mu_{y^{'}}$ (corresponding to the minor axis); with $d$ the distance in kpc and $i$ the inclination of the rotation axis. A detailed comparison between line-of-sight datasets and proper motion datasets is beyond the scope of this work; however, we note that along the axis perpendicular to the rotation axis (i.e. the major axis), $\mu_{y^{'}}=\mu_t$. The relation must also be valid at the peak of the rotation curve (which is located at a similar position both for the line-of-sight and for the proper motion rotation curves). Therefore, we can use the values of $V/\sigma$ from this work and the ones for the line-of-sight from the literature to obtain a quantitative estimate of the inclination angle, applying equation \ref{eq:i_vlos}. We obtain $i\simeq26^\circ$ for NGC104 and $i\simeq44^\circ$ for NGC5139 (fully consistent with \citealp{Bellini2017} and \citealp{vandeVen2006}, respectively); $i\simeq49^\circ$ for NGC4372, $i\simeq38^\circ$ for NGC5904 and $i\simeq29^\circ$ for NGC7078, consistent with the qualitative discussion presented in the paragraph above. The results are reported in Table \ref{tab:inclination}.

An alternative way to measure the inclination of the rotation axis consists of exploiting the spatial information contained in the two-dimensional rotation maps presented in Section \ref{sec:rot_maps}. Assuming that all stars in an axisymmetric GC rotate around the symmetry axis with velocity $v_{rot}$, then $v_{rot}$ can be written as the vectorial sum of the line-of-sight rotation component and the plane-of-the-sky tangential velocity component, $v^2_{rot}=v^2_t+v^2_{los}$. Limiting the analysis on a slit along the minor axis $v^{min}_{los}=0$; while along the major axis, $v^{maj}_{los}=v_{rot}\sin i$, with $i$ the inclination of the rotation axis. It follows that 
\begin{equation}
\begin{aligned}
v^{min}_t &= v_{rot}\\
v^{maj}_t &= v_{rot}\cos i.
\end{aligned}
\label{eq:i_maps}
\end{equation}
Therefore, we can obtain an estimate of the inclination by comparing the mean tangential proper motion along the minor and the major axis. Using the values of the position angle of the rotation axis available from line-of-sight studies (see Table \ref{tab:inclination} and Figures \ref{fig:rot_curve1}, \ref{fig:rot_curve2} and \ref{fig:rot_curve3}) we rotate the clusters so that the y axis coincides with the rotation axis. We then consider all the stars within a slit with size $\Delta$ along the major and minor axis and calculate $\mu_t^{maj}$ and $\mu_t^{min}$ using the likelihood function given in equation \ref{eq:loglike}. Comparing the values of $\mu_t^{maj}$ and $\mu_t^{min}$ through equation \ref{eq:i_maps} we estimate the inclination angle for 8 GCs. We report the results in Table \ref{tab:inclination}. We note that the values of inclination obtained using the two different methods outlined in this section (equations \ref{eq:i_vlos} and \ref{eq:i_maps}) are fully consistent with each other.

\section{Conclusions}
We exploit the proper motion sample from \textit{Gaia} DR2 to conduct a systematic search for rotation signatures on the plane-of-the-sky for a large sample of MW GCs. First we select 51 clusters for which a signature of rotation of $\simeq1$ km s$^{-1}$ is likely to be measurable given the number of measured stars and their distance. For these clusters we select \textit{Gaia} DR2 data within three half-light radii of their centres and perform a cleaning of the sample to select probable cluster members and discard contaminants. For each cluster we measure the mean tangential proper motion component within the selected field-of-view and consider as a rotation detection the cases in which the mean velocity is significantly different than zero at least at a 3-sigma confidence level. With this procedure we identify 11 rotating clusters: NGC104, NGC5139, NGC5272, NGC5904, NGC6656, NGC6752, NGC6809, NGC7078 (for which rotation was reported by \citealp{Helmi2018}) and additionally NGC4372, NGC6273 and NGC7089. Additionally, we reported 11 GCs with rotation detection at a 2-sigma level (NGC3201, NGC5286, NGC6121, NGC6205, NGC6254, NGC6266, NGC6397, NGC6402, NGC6539, NGC6541, NGC6779) and  identify these GCs as the ideal targets for follow-up internal rotation studies.

For each of the rotating clusters we construct proper motion rotation profiles and two-dimensional velocity maps. The rotation in the tangential component of our clusters is characterized by a differential rotation structure, with low rotation amplitudes near the centre, a peak in the region around 0.5-2 $R_h$, and an outward decrease. The values of the rotation peak span a range between $\simeq0.7-6$ km s$^{-1}$. For consistency, we check the profiles for the radial component of the proper motions and find them to be consistent with the expected perspective expansion/contraction due to the clusters' bulk motions along the line-of-sight.

In order to quantify the relevance of rotation with respect to random motion we measure the $V/\sigma$ parameter on the plane-of-the-sky, finding values between $V/\sigma\simeq0.08-0.51$. We report evidence for a mild correlation between $V/\sigma$ and the relaxation time of a cluster, indicating that dynamically relaxed clusters have more efficiently dissipated their internal angular momentum, due to their long-term dynamical evolution. Moreover, we find a correlation between the $V/\sigma$ and the metallicity of a cluster, with metal-rich clusters exhibiting higher values of rotation. Both of these correlations convey important pieces of information for the early phases of cluster formation and subsequent dynamical evolution. However, we notice that a thorough interpretation of these results would require the study of the three-dimensional intrinsic angular momentum, in order to avoid projection effects.

We show in this work that the kinematic information from our proper motion analysis can be used to estimate the three-dimensional structure of the clusters. We are able to constrain the inclination of the rotation axis for 8 GCs, using two different methods. We confirm previous literature studies indicating that NGC104 and NGC5139 have an inclination angle $i\simeq30^{\circ}$ and $i\simeq45^{\circ}$, respectively. For NGC4372, NGC5904, NGC6656, NGC6752, NGC7078 and NGC7089 we provide the inclination measurement for the first time.

Our analysis reveals the strength of \textit{Gaia} proper motions for internal dynamical studies of GCs and sets the basis for future comparisons with line-of-sight kinematics. Finally, our work demonstrates the fundamental role of angular momentum in todays GCs and emphasizes the need for comprehensive three-dimensional dynamical studies to shed light onto cluster dynamical evolution, cluster formation and proto-GCs properties.


\section*{Acknowledgments}
We wish to thank the referee for useful comments that helped improve the quality of the paper.
PB acknowledges financial support from a CITA National Fellowship.
This work has made use of data from the European Space Agency (ESA)
mission {\it Gaia} (\url{https://www.cosmos.esa.int/gaia}), processed by
the {\it Gaia} Data Processing and Analysis Consortium (DPAC,
\url{https://www.cosmos.esa.int/web/gaia/dpac/consortium}). Funding
for the DPAC has been provided by national institutions, in particular
the institutions participating in the {\it Gaia} Multilateral Agreement.
This work made use of the PyGaia package provided by the Gaia Project Scientist Support Team and the Gaia Data Processing and Analysis Consortium (\url{https://github.com/agabrown/PyGaia}).
This project is part of the HSTPROMO (High-resolution Space Telescope PROper MOtion) Collaboration\footnote{\url{http://www.stsci.edu/~marel/hstpromo.html}}, a set of projects aimed at improving our dynamical understanding of stars, clusters and galaxies in the nearby Universe through measurement and interpretation of proper motions from \textit{HST}, \textit{Gaia}, and other space observatories. We thank the collaboration members for the sharing of their ideas and software.

\bibliographystyle{mnras} 
\bibliography{biblio} 

\end{document}